\documentclass{emulateapj}
\usepackage{natbib}
\usepackage{graphicx}
\usepackage{epstopdf}
\DeclareMathSymbol{\varOmega}{\mathord}{letters}{"0A}
\DeclareMathSymbol{\varSigma}{\mathord}{letters}{"06}

\DeclareMathSymbol{\varPsi}{\mathord}{letters}{"09}

\newcommand{\Eq}[1]{Equation\,(\ref{#1})}
\newcommand{\Eqs}[2]{Equations (\ref{#1}) and~(\ref{#2})}

\newcommand{\Fig}[1]{Figure~\ref{#1}}
\newcommand{\Figs}[2]{Figures~\ref{#1} and \ref{#2}}

\newcommand{\eqfrac}[2]{\left(\frac{#1}{#2}\right)}
\newcommand{\icarus}{\rm{Icarus}}

\newcommand{\ltsim}{\protect\raisebox{-0.5ex}{$\:\stackrel{\textstyle <}{\sim}\:$}}
\newcommand{\gtsim}{\protect\raisebox{-0.5ex}{$\:\stackrel{\textstyle >}{\sim}\:$}}

\newcommand{\ddx}{\frac{{d}}{{d} x}}

 \newcommand{\Mdot}{\dot{M}}
\newcommand{\Msun}{M_{\odot}} 
\newcommand{\Lsun}{L_{\odot}}

\newcommand{\lsim}{\mathrel{\rlap{\lower4pt\hbox{\hskip1pt$\sim$}}
    \raise1pt\hbox{$<$}}}                % less than or approx. symbol
\newcommand{\gsim}{\mathrel{\rlap{\lower4pt\hbox{\hskip1pt$\sim$}}
    \raise1pt\hbox{$>$}}}                % greater than or approx. symbol

\begin{document}

%\title{When Unstable Regimes Collapse: History, Long-Term Instability, \\And the cooling criterion for self-gravitating disks}
%\title{Fragment Production and Survival in Strongly Irradiated Disks: A Comprehensive Cooling Criterion}
\title{Fragment Production and Survival in Irradiated Disks: \\A Comprehensive Cooling Criterion}
\begin{abstract}

Accretion disks that become gravitationally unstable can fragment into stellar or sub-stellar companions. The formation and survival of these fragments depends on the precarious balance between self-gravity, internal pressure, tidal shearing, and rotation. Disk fragmentation depends on two key factors (1) whether the disk can get to the fragmentation boundary of Q=1, and (2) whether fragments can survive for many orbital periods.  Previous work suggests that to reach Q=1, and have fragments survive, a disk must cool on an orbital timescale. Here we show that disks heated primarily by external irradiation always satisfy the standard cooling time criterion. Thus even though irradiation heats disks, and makes them more stable in general, once they reach the fragmentation boundary, they fragment more easily. We derive a new cooling criterion that determines fragment survival, and calculate a pressure modified Hill radius, which sets the maximum size of pressure-supported objects in a Keplerian disk. We conclude that fragmentation in protostellar disks might occur at slightly smaller radii than previously thought, and recommend tests for future simulations that will better predict the outcome of fragmentation in real disks.

\end{abstract}

\author{Kaitlin M. Kratter\footnote{Email: kkratter@cfa.harvard.edu}  and Ruth Murray-Clay}
\affil{Harvard-Smithsonian Center for Astrophysics, 60 Garden St., MS-51, Cambridge, MA, USA, 02138}
\maketitle 

\section{Introduction}

The past two decades have witnessed growing interest in the conditions under which gravitationally unstable disks fragment to produce bound objects. These fragments may grow into massive stars or black holes within AGN disks, or small stellar and substellar companions in protostellar disks. In either context, we can divide the process into two steps. First, the disk must become linearly gravitationally unstable, and secondly, once formed, the fragments must be able to survive the various disruptive forces in the disk. 

The first step, linear gravitational instability (GI),  requires that the disk be Toomre unstable: $ Q= c_s \Omega / \pi G \Sigma \leq Q_0 \approx 1$ \citep{Toom1964,Saf1960},  where $c_s = \sqrt{kT/\mu}$ is the isothermal sound speed of the gas with mean particle weight $\mu$ and temperature $T$, $G$ is the gravitational constant, $\Sigma$ is the disk surface density, $k$ is the Boltzmann constant,  and $\Omega$ is the orbital angular frequency. 

The second stage is more uncertain. Three primary forces compete to destroy newly born fragments: internal gas pressure, tidal forces from the central object  (shearing), and internal rotation. Fragments can also destroy each other through mutual interactions.

Whether the disk can be driven to the fragmentation boundary, which we call $Q=Q_0$, and whether fragments can survive, must be considered separately.  Both requirements place limits on the cooling properties of disks. Moreover, the answer to both of these questions depends on whether the disk is actively heated (i.e. by dissipation of accretion energy through an effective turbulent $\alpha$ viscosity), or passively heated (i.e. be external irradiation).  Both processes operate in most disks, but it matters which is the dominant energy source. Here we consider the passive (irradiated) case in detail, as it has been neglected in previous theoretical studies. { Throughout the text we use the term viscosity to describe any process that contributes to both transport of angular momentum and local energy dissipation.}

The role of cooling in fragmentation has been discussed in the literature in two ways, which we dub the Viscous Criterion \citep{Pringle81,Gam2001}  and the Collisional Criterion \citep{1989ApJ...341..685S}. The Viscous Criterion describes whether an actively-heated disk can rid itself of the energy generated by GI-driven accretion on the disk's dynamical timescale, while the Collisional Criterion considers whether newborn fragments will collide with each other on this timescale. While the Viscous Criterion addresses the first question: (1)  Can a disk be driven to $Q=Q_0$ so that initial collapse begins, the Collisional Criterion addresses the second: (2) Can a collapsing fragment survive? 

Here we provide a more complete picture of the necessary disk conditions for fragmentation and fragment survival.  We begin with question (1), emphasizing the crucial role of disk history in determining whether a disk can be driven to collapse.  The Viscous Criterion answers this question solely for viscously-heated disks, and we provide a complementary criterion for irradiated disks, which we show automatically satisfy the Viscous Criterion.  We then turn to question (2).  We find that the Collisional Criterion does not fully capture the role of cooling in fragment survival, because it does not account for the contraction of fragments.  We propose a more general cooling criterion which governs fragment survival in all disks. This ``Stalling Criterion" first determines whether fragments collapse on the free-fall timescale, or if they quickly become pressure supported.  If a fragment becomes pressure supported, stalling collapse, it must be smaller than a pressure-modified Hill radius to avoid tidal disruption.  Fragments that contract to small enough sizes before stalling can survive while slowly undergoing Kelvin-Helmholz contraction.  

\begin{figure*}
\begin{centering}
\plotone{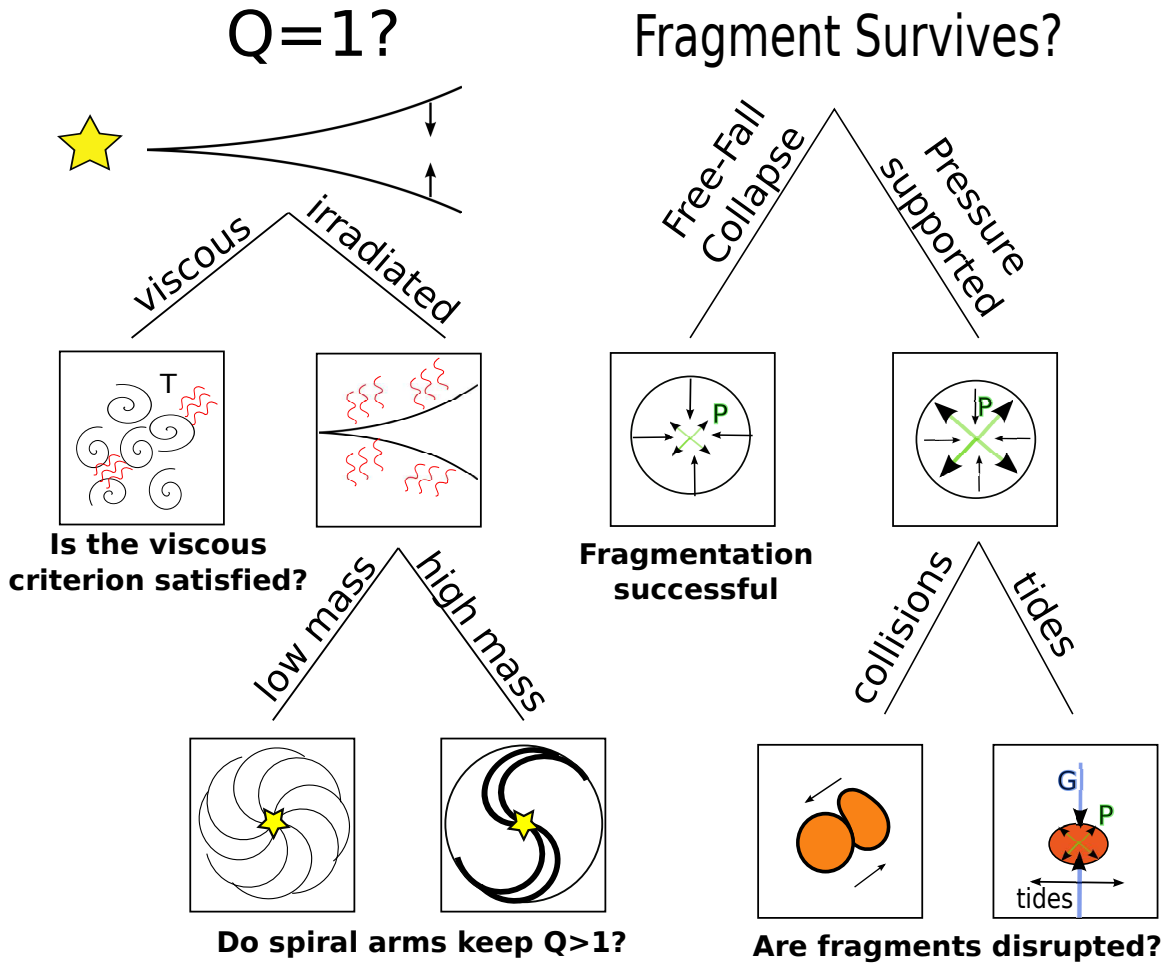}
\caption{Flow chart illustrating the relevant physics and criteria for determining disk fragmentation. First, the disk much reach $Q=1$. In viscously heated disks this is determined by the Viscous Criterion (\S\ref{sec-viscdrive}), and in irradiated disks by whether global modes can process mass as fast as it infalls on to the disk (\S\ref{sec-irraddrive}). If $Q \rightarrow 1$, then one must determine if fragments survive. Collapsing in free fall likely guarantees survival (\S\ref{KHC}). Fragments which do not collapse quickly must avoid collisional (\S\ref{sec-ccrit}) or tidal disruption by shrinking below a critical size, $R_{\rm H+p}$ (\S \ref{sec-KHcoll}).}
\label{fig-3crit}
\end{centering}
\end{figure*}

Fragment survival depends both on the cooling time and the gas equation of state. The Stalling Criterion is applicable in both viscously-heated and irradiated disks. \Fig{fig-3crit} illustrates how the different cooling criteria we describe determine whether disks fragment, and whether their fragments survive.

We begin with brief definitions of the disk cooling time and related terms in Section \ref{coolingtimes}. We then distinguish quantitatively between disks heated via external irradiation or internal viscous dissipation in Section \ref{sec-irradmean} and describe the run-up to instability in each case (Sections \ref{sec-viscdrive} and \ref{sec-irraddrive}). We introduce our new cooling criterion in Section \ref{sec-KHC}. We show that for arbitrary cooling time, the Stalling Criterion sets a critical size scale ($R_{\rm H+p}$) below which a fragment can survive if pressure supported.  Whether or not a fragment can reach this critical radius depends on the disk equation of state.  We review other means for disruption of fragments in Section \ref{caveats}.

 We discuss the implications of these processes for protostellar and AGN disks in Section \ref{sec-irrad}, then comment briefly on recent simulations of viscously-heated disks (Section \ref{sec-visc}).   Finally, we provide suggestions for future simulations of irradiated disks and conclude in Section \ref{discussion}.  
 
 \section{The Definition of the Cooling Time}\label{coolingtimes}
For clarity we begin with a few definitions. In general, we consider a disk with cooling time 
\begin{equation} \label{eq-taudef}
\tau_c = \beta \Omega^{-1}
\end{equation}
where $\tau_c$ is defined by the following term in an energy equation:
\begin{equation}\label{eqn-coolterm}
\left.\frac{\rm {du}}{\rm dt}\right|_{\rm cool} = -\frac{u}{\tau_c},
\end{equation} 
$\beta$ is a constant, and $u$ is the internal energy per unit mass. 
The use of the $\beta$ cooling formalism is motivated by the fact that the orbital time is the important dynamical timescale, and allows for a cancellation in the energy equation. The result will be a dimensionless cooling criterion.

In reality, $\tau_c$ is a function of the physical  properties of a given disk, and thus varies in time and location. \cite{KMCY10} calculate the cooling time for a thermal perturbation from the midplane to be\footnote{In strongly viscously-heated disks, an order unity correction factor must be applied to the coefficient of Equation (\ref{eqn-tcooldef}) if the opacity of disk material depends on temperature \citep[see][]{KMCY10}.  Including this factor does not change our results.}:
\begin{equation} \label{eqn-tcooldef}
t_{\rm cool } =\frac{3}{32}\frac{\gamma}{\gamma-1}\frac{\Sigma c_s^2}{\sigma T^4}\tau_R
\end{equation}
where $\tau_R$ is the optical depth due to the Rosseland mean opacity, $\tau_R= \kappa \Sigma/2$ , $\gamma$ is the adiabatic index, $\sigma$ is the Stefan-Boltzmann constant, and we have assumed that the disk is optically thick, which is valid for $Q=1$ disks. We use \Eq{eqn-tcooldef} to translate between dimensionless cooling times and physical disk conditions, however we quote our analytic expressions in terms of $\tau_c$ for generality and convenience.

With these definitions in hand, we now describe when astrophysically relevant disks are susceptible to fragmentation.

\section{ Driving Disks to Linear Instability}
The required initial conditions for creating gravitationally unstable disks depend on the dominant energy source in the disk. As alluded to above, passively heated disks behave differently from actively heated ones. First we clarify this distinction, and then review the required conditions for disks in each class to become linearly unstable to GI.

\subsection{Irradiated vs Viscously Heated Disks}\label{sec-irradmean}
What does it mean for a disk to be ``dominated" by irradiation?  Qualitatively it means that the primary source of the disk's thermal energy is external irradiation rather than dissipation of accretion energy (which we compute as an artificial viscosity). In other words, the temperature to which the disk { midplane} is heated by external irradiation is higher than the equilibrium temperature that the disk could achieve by balancing accretional heating and cooling via thermal emission. As a result, irradiated disks will be roughly vertically isothermal. We shall show an example of this later in \Figs{fig-proto}{fig-both}. 

 We choose this terminology over the more common, but less informative, distinction between passive (irradiated), and active (viscously heated), primarily because irradiated disks can have ongoing accretion -- they are not entirely ``passive."

Quantitatively, the boundary between an irradiated and viscously heated disk can be defined by balancing radiative cooling from the midplane with heating via dissipation of accretion energy.  We quantify this dissipation in the standard manner, by employing an effective viscosity, $\nu = \alpha c_s^2/\Omega$, described by the parameter $\alpha \leq 1$. The { midplane} temperature, $T$, generated by dissipation of energy due to accretion of a mass flux 
\begin{equation}\label{eqn-mdot}
\dot M = 3\pi\Sigma\nu
\end{equation} 
is given by:\begin{equation}\label{eqn-accheat}
\frac{1}{\tau_{R}}\sigma T^4 = \frac{3}{8\pi}\Mdot \Omega^2 = \frac{9}{8}\alpha c_s^2\Sigma\Omega 
\end{equation}
{in the optically thick case, relevant for $Q=1$ disks}.
Inserting \Eq{eqn-accheat} into \Eq{eqn-tcooldef} yields 
\begin{equation}\label{eqn-equilibtcool}
t_{\rm cool} = \frac{1}{12\alpha}\frac{\gamma}{\gamma-1}\Omega^{-1} \;. 
\end{equation}
{Equation (\ref{eqn-equilibtcool}) expresses the conditions for thermal equilibrium of a disk  heated only by viscous dissipation in terms of the cooling time as defined in \Eq{eqn-tcooldef}}.
Irradiated disks have cooling times less than this critical value:
\begin{equation}\label{eqn-irraddom}
t_{\rm cool} < \frac{1}{12\alpha}\frac{\gamma}{\gamma-1}\Omega^{-1} \;, 
\end{equation}
because they are heated to higher temperatures by irradiation than by dissipation of accretion energy. This critical disk temperature, in terms of density and $\alpha$ is:
\begin{equation} \label{eq-thickcool}
T > \left(\frac{9}{8}\frac{\alpha\Sigma}{\sigma}\frac{k}{\mu}{\tau_R}\Omega\right)^{1/3}.
\end{equation}

If stellar irradiation maintains the disk (or some region of it) above the temperature in \Eq{eq-thickcool}, it is irradiation dominated. {Typically the outer regions of disks are irradiation dominated, while in their inner regions accretional heating is more important than irradiation}. This radial dependence is due to the liberation of more energy as material in the disk is lowered deeper into the potential well of the central object.

To evaluate whether a disk in steady-state satisfies \Eq{eq-thickcool} at a particular disk location, one must set $\alpha$ to the value required to process the incoming material either from larger radii in the disk or the background environment (e.g. a protostellar core). Of course, if the $\alpha$ required to process the material is larger than that provided by any means of angular momentum transport, a steady state is not possible.  We note that over time, a particular disk radius may transition from irradiation-dominated to viscously heated or vice versa as the disk evolves.

The boundary between viscously heated and irradiated disks is independent of the mechanism for angular momentum transport, so long as it can be quantified by an $\alpha$. 
In this work we are concerned with transport due to GI.  Unlike transport due to the magneto-rotational instability (MRI) \citep{BH94}, which is often modeled using a single value for $\alpha$, GI-induced transport cannot be described by a fixed $\alpha$.  Instead, $\alpha$ is a function of disk parameters (in particular $Q$) and should be thought of as a response to the self-gravitating state of the disk \citep{1987MNRAS.225..607L}.  We return to this point in Section \ref{sec-viscdrive}.

We now answer the first question posed above, and in \Fig{fig-3crit}: how do disks reach the boundary of linear instability? 

\subsection{Driving a Viscously-Heated Disk to Instability -- the Viscous Criterion}\label{sec-viscdrive}

In order to fragment, a disk must first reach $Q=Q_0\sim1$, allowing linear collapse to proceed.  Though dramatic events may sometimes drive disks into gravitationally unstable configurations on short timescales, typical disks likely evolve on long enough timescales that $Q=1$ must be approached gradually if it is approached at all.  

First, consider disks which are heated only through accretion energy. In the absence of any external heat source (e.g. irradiation), a disk can only be driven to fragment by  rapid infall \citep{ML2005}.  In the case of a protostellar disk, for example, this infall comes from the protostellar cloud and its rate is set by conditions external to the disk.\footnote{Evolution to $Q=Q_0$ might also occur in protostellar disks after infall from the protostellar cloud ceases if the efficiency of surface accretion due to the MRI varies as a function of distance from the star (\citealt{ZhuHartGam10}, Perez-Becker \& Chiang, in prep).  The following still applies, however, with surface accretion from outer regions of the disk adopting the roll of ``infall."} We discuss in the next section how irradiation changes this picture, {and in the appendix we derive explicitly the evolution of $Q$ with $d\Mdot/dt$}.

As long as infall rates are not too large, a $Q\gtrsim 1$ disk heated solely by viscous dissipation relaxes into an equilibrium state known as gravitoturbulence.  Gravitoturbulence results from a feedback loop which may be understood as follows. When $Q\sim 1$, if radiative cooling exceeds viscous heating, the disk will cool and get closer to $Q=1$. As it gets closer to $Q=1$, the onset of gravitational collapse leads to turbulence and accretion, which heats the disk and generates extra pressure support to counteract gravitational collapse.  If the heating rate were to exceed the cooling rate such that $Q\gg1$, then the $\alpha$ generated by GI would decrease, the heating rate would decline, and the disk would again cool down. This feedback loop is gravitoturbulence. 

In gravitoturbulent equilibrium, viscous heating and radiative cooling balance such that
\begin{equation}\label{eq-tequil}
\beta = \frac{4}{9\gamma(\gamma-1)}\frac{1}{\alpha}
\end{equation}
\citep{Gam2001}.
Here, $\beta$, defined in \Eq{eq-taudef}, represents the cooling rate in terms of orbital periods.  
One may derive Equation (\ref{eq-tequil}) by equating the cooling rate per unit mass $u/\tau_c$ (Equation \ref{eqn-coolterm}) with the viscous heating rate per unit mass, $(9/4)\alpha c_s^2\Omega$ (c.f. Equation \ref{eqn-accheat}, remembering to include two disk faces).  We have used $u = c_s^2/(\gamma - 1)$, appropriate for an ideal gas.\footnote{The extra factor of $\gamma$ in the denominator of Equation (\ref{eq-tcoolalpha}), included for consistency with \citet{Gam2001}, comes from that paper's use of the adiabatic sound speed rather than the isothermal sound speed in the definition of $Q$.}.

However, the equilibrium reflected in Equation (\ref{eq-tequil}) cannot be achieved for arbitrarily high mass-infall rates.  For GI-driven accretion, $\alpha$ is a function of disk properties (in particular of $Q$),  dictated by the strength of self-gravity in the disk.  Numerical simulations demonstrate that there exists a maximum transport rate provided by gravitational instability 
\citep{Gam2001}
and hence a maximum $\alpha <1$, which we call its saturation value $\alpha_{\rm sat}$.
  
The precise value of $\alpha_{\rm sat}$ as a function of disk properties, and the function $\alpha(Q)$ remain uncertain (though see, \citealt{KMK08, 2010NewA...15...24V}). { We emphasize that $\alpha$ is not a free parameter, and neither is the temperature of the disk in equilibrium.  Whether or not a disk can restablize itself at ever higher $\Mdot$ depends on both $\alpha(Q)$ and on the temperature dependence of the opacity. Unless $\kappa \propto T^{> 2}$, increasing $\Mdot$ always drives $Q$ down. Depending on the functional form of $\alpha(Q)$, a very steep temperature dependence may also cause $Q$ to decline with increasing $\Mdot$. The disks that we consider here typically have $\kappa \propto T^{\leq2}$ \citep{2003A&A...410..611S}, but we explore this dependence fully in Appendix A. Thus in general, we expect rapid accretion to drive disks toward instability, and in some cases fragmentation.}

The distinction between a disk that can maintain gravitoturbulence and one that fragments is encompassed in the Viscous Criterion:
\begin{equation}\label{eq-tcoolalpha}
\beta \leq \frac{4}{9\gamma(\gamma-1)}\frac{1}{\alpha_{\rm sat}}.
\end{equation}
When Equation (\ref{eq-tcoolalpha}) is satisfied, cooling exceeds viscous heating and the disk is unable to reverse this imbalance by increasing its accretion rate, and hence its heating rate.  { Thus material is being added into the disk faster than it can be drained to smaller radii, which causes $\Sigma$ to increase proportionally faster than $T^{1/2}$, $Q$ to decline, and fragmentation to proceed.}
We note that accretional energy must be deposited in the disk as material falls toward the star, independent of the physical processes involved in dissipation of that energy.  Hence, the Viscous Criterion is robust unless dissipation of accretion energy is dramatically non-local.    

Note that most simulations of gravitoturbulence involve isolated disks, i.e. not undergoing infall. These disks fragment because they are set up {\emph {out of equilibrium}} \citep{Lod04,LodRi05}. Cooling exceeds heating by fiat, and thus the disk moves towards $Q=1$. In reality, the cooling rate is set by the thermal physics of the disk, and is not an independent parameter.

Of course, as we addressed above, many disks have two heating sources, accretion energy and external irradiation. 
In fact, the Viscous Criterion is nearly identical to the boundary between these two regimes in \Eq{eqn-irraddom}, modulo different numerical coefficients.
 The difference in coefficients derives from the fact that here we are concerned with the timescale for the disk to radiate away its internal energy, and therefore cool off.  To distinguish between viscous and irradiated disks we instead consider whether a temperature perturbation in the midplane heats the disk or is radiated away  (see the appendix of \citealt{KMCY10}). 
 A disk satisfying the Viscous Criterion can avoid fragmentation only if it is irradiation dominated, because the external heat source alters the feedback loop -- the temperature, and thus cooling rate -- are no longer controlled by dissipation, but instead by the radiation field. 
   We address how disks heated by irradiation are driven towards instability below.

\subsection{Driving an Irradiated Disk to Instability}\label{sec-irraddrive}

What is required for an irradiated disk to be driven to $Q = Q_0$? Again,  infalling material is required to drive the disk unstable.  However, we suggest that for very low mass disks, the critical infall rate may be lower than previously thought. 

   Unlike their viscously heated counterparts,  irradiated disks will not become gravitoturbulent as infall adds mass and decreases $Q$. Gravitoturbulence relies on a feedback loop between $Q$ and disk temperature, but the temperature of an irradiation-dominated disk is independent of $Q$, modulo small perturbations.  For simplicity, for the remainder of this section, we only consider disks which remain irradiation dominated at all accretion rates.

To understand why irradiated disks cannot maintain gravitoturbulence, consider the following hypothetical disk, with $Q=1$ around a $1.5\Msun$ star. Imagine that as the disk approaches $Q=1$, the effective $\alpha$ generated by GI-driven accretion increases from 0.1 to $\alpha_{\rm sat} = 1.0$.  For an order of magnitude increase in $\alpha$, the equilibrium temperature for a disk heated through viscous dissipation rises from $5-20$K at 100 AU. However, the equilibrium temperature of the disk including external irradiation from a young A star increases from $47-50$K. Thus the fractional contribution of GI to the total pressure support of the disk ($\propto T$) is negligible.  Now, we add more and more material to the disk. Even if the disk processes material at this maximum rate, and dissipates energy at this maximum rate,  the disk cannot reheat itself via turbulence. External irradiation, which initially prevents the disk from collapsing, cannot adjust the disk temperature to avoid the $Q=1$ threshold.  Once $Q$ crosses this threshold, fragmentation will proceed. 

However, irradiation-dominated disks are not entirely unaffected by self-gravity as infalling material drives them toward $Q=1$.  Irradiated disks (or indeed disks with an isothermal equation of state, which behave similarly) respond to decreasing $Q$ by generating gravitational torques in spiral arms \citep{Boley2007,KMKK10}.  These torques can, in turn, drive rapid accretion through the disk.

The properties of spiral arm transport depend on the ratio of the disk mass \citep{ARS89}, $M_{\rm disk} = \pi r_{\rm out}^2 \Sigma$, to the central mass, $M_*$.  Here $r_{\rm out}$ is the outer radius of the disk, where most of the mass resides.  For a $Q=1$ disk, $M_{\rm disk}/M_* = H/r_{\rm out}$, 
where $H = c_s/\Omega$ is the disk scale height, so the disk-to-central object mass ratio is equivalent to the disk aspect ratio.  Spiral arms that form in thin and thick disks have different properties.

Numerical simulations indicate that high-mass (thick) disks can process mass infall rates up to  $\dot M \sim c_s^3/G$ \citep{KMKK10, OKMKK10}, while at higher rates they fragment.  Lower infall rates do not generate fragmentation because these disks become globally gravitationally unstable; the resulting spiral arms transport material efficiently, keeping $Q>Q_0$.  In the language of Section \ref{sec-viscdrive}, $\dot M \gtrsim c_s^3/G$ corresponds to driving the disk with an infall rate having $\alpha \gtrsim 1$.  Thus thick, irradiated disks, like viscously-heated disks, must be forced with large infall rates in order to fragment.

However, if the disk is low in mass (thin), has inefficient transport by spiral arms, and is very near $Q=1$, a modest infall rate could tip it over the fragmentation threshold faster than it can process even the modest amount of mass it receives. For this to occur, other mechanisms of angular momentum transport such as the MRI must not be able to process the infalling material. 
Whether low-mass (thin), irradiated disks can avoid effective angular momentum transport by spiral arms remains uncertain.
Simulations suggest that strong global spiral arm transport may require $M_{\rm disk}/M_* \gtrsim 0.1$ \citep{Lod04}.  

Since the actual criterion for global mode transport is unknown, it is difficult to provide a precise criterion:  future numerical work can delineate the exact conditions for this regime, and indeed whether it exists at all. This work will need to be conducted with great care; N-body and SPH simulations which show spontaneous instability at $Q >1$ may suffer from Swing amplification of Poisson noise (D'Onghia et al 2011, in prep). Current state of the art simulations \citep{Boley2007,2008ApJ...673.1138C} do not yet explore the full parameter space required to answer these questions. For earlier attempts to quantify this behavior see \cite{1996ApJ...456..279L,1996ApJ...460..855L,1997ApJ...477..410L}

Can nature produce a $Q\approx 1$ disk that remains irradiation-dominated, unaffected by the MRI, and low enough in mass to avoid global instability?  We show in \S\ref{sec-irrad} that it is difficult, but not impossible to achieve $Q=1$ and maintain a sufficiently low mass disk to suppress  spiral arm formation. 

Thus we have answered the first of our two questions and shown how both classes of disks can reach the instability threshold. In summary, to fragment, viscously-heated disks 
must be driven by mass infall at a rate corresponding to $\alpha > \alpha_{\rm sat}$.  Disks that are irradiation-dominated (and thus satisfy the Viscous Criterion) might be driven to fragment at lower accretion rates, particularly when such disks have low masses relative to their host stars (i.e., are thin). 
This possibility is interesting because such disks could generate fragments that remain near their initial masses rather than experiencing substantial growth.  Numerical simulations of global mode formation in low-mass, irradiated disks will be required to evaluate this possibility.  Other angular momentum transport mechanisms such as the MRI will prevent fragmentation from occurring for arbitrarily small infall rates.

Can fragments survive once the instability threshold is reached? We now consider the cooling criteria related to fragment survival: the Stalling Criterion, and the Collisional Criterion.

\section{Determining Fragment Survival} \label{sec-KHC}
Long term fragment survival requires that the fragment's own self-gravity wins out over internal pressure, rotation, tidal forces from the central object, and interactions with fellow fragments. We consider the importance of the first three forces through the Stalling Criterion, then argue that in all regimes of astrophysical interest, fragments which satisfy the Stalling Criterion are also immune to collisional disruption.

The Stalling Criterion concerns the evolution of an already-formed fragment---in other words, it constrains the non-linear outcome of successful linear collapse. 
If a disk can reach the fragmentation threshold, 
two possible fates befall a newly formed fragment.   If the fragment cools on its own free-fall timescale,  it can contract quickly to a length-scale that is small compared to its Hill radius.  Conversely, if cooling is slow compared to the free-fall timescale, then the fragment's collapse will be stalled by pressure support, and it will begin to collapse on a modified Kelvin-Helmholtz timescale, of order its cooling time.

We expect a fragment that collapses on the free-fall timescale to survive, although we discuss other mechanisms for disruption in \S\ref{caveats}.  
Will a fragment whose collapse is stalled by pressure support on the disk dynamical time survive? It depends on the radius at which pressure halts collapse.  
To survive, a pressure supported fragment must be smaller than 
a critical size scale, which we call the pressure-modified Hill radius, $R_{\rm H+p}$.  
Once a fragment has shrunk to a size smaller than $R_{\rm H+p}$, it can survive regardless of its cooling rate (modulo other mechanisms for disruption discussed in \S \ref{caveats}).  

Shrinking quickly to a fraction of a Hill radius may prove impossible for some slowly cooling fragments, depending on the gas equation of state and on the initial density perturbation seeding collapse.  We calculate the requirements for collapse on a free-fall time in \S\ref{KHC}, derive $R_{\rm H+p}$ in \S\ref{sec-KHcoll}, and estimate the initial conditions required for fragment survival in \S\ref{stalling-rad}.  

We comment that the Stalling Criterion has been overlooked primarily because gravitoturbulence is thought to generate a large $\alpha_{\rm sat}$. If $\alpha_{\rm sat}$ is large, a disk which cools quickly enough to be either irradiation dominated or satisfy the Viscous Criterion (see Equations \ref{eqn-irraddom} and \ref{eq-tcoolalpha}) will usually cool quickly enough to ensure free-fall collapse of the fragment. However the physics controlling fragment survival is distinct from that controlling the approach to the $Q=1$ boundary, and for smaller $\alpha_{\rm sat}$, or  disks with sharp opacity jumps, both criteria should be considered.

\subsection{ Survival via Free-Fall Collapse}\label{KHC}
 Consider a disk that has arrived near enough to $Q=1$ to produce perturbations that can seed collapse. In an isothermal disk, once a perturbation exists with $Q<1$, collapse will occur on the free fall time scale. For non-isothermal collapse we must examine the energy budget of the collapsing region as self-gravity tries to make it contract. 
 
 We neglect internal rotation in the following analysis as it is difficult to estimate its evolution in time analytically. We comment on the importance of rotation in \S\ref{caveats}.  By neglecting rotation we provide optimistic limits for collapse and thus fragment survival. 
  
When $Q \sim 1$ and linear collapse has begun,  self-gravity, pressure, and tidal forces from the central object exert comparable forces on the fragment, but self-gravity is currently winning.  As collapse proceeds and the radius, $x$ of the fragment gets smaller, the ratio of the self-gravitational force to the tidal force increases as $x^{-3}$, assuming that the enclosed mass in the fragment stays constant.  However, the ratio of self-gravitational force to pressure force may increase or decrease. We define
the virial ratio $B$ in a clump with density  $\rho$:
 \begin{equation}
B = \frac{G M_{enc}\rho /x^2}{P/x},
\end{equation}
where $M_{\rm enc} = 4/3\pi \rho x^3$, and $P$ is the internal pressure. We approximate the pressure gradient as $P/x$ since the fragment is only one pressure scale height in extent.\footnote{Including tidal gravity in B causes a small change in the derived coefficient for the free-fall regime. We exclude it here for simplicity. }
If $B$ increases with time, then self-gravity will always dominate and free fall collapse is ensured. If $B$  decreases, pressure will eventually stall collapse.

For small enough $\beta$ (fast cooling), $B$ initially increases as collapse proceeds.  In other words, 
\begin{equation}\label{eqn-Breq}
-\ddx(B) > 0   \;,
\end{equation}
where the negative sign comes from the fragment radius, $x$, decreasing as the object collapses.
Fragments collapse in free fall when $\beta$ is less than a critical value, which we now calculate.

Holding fragment mass, $M_{\rm enc}$, fixed and substituting in for an ideal gas with pressure $P= (\gamma-1)\rho u$, Equation (\ref{eqn-Breq}) yields the following inequality:
\begin{equation}\label{eq-simpledx}
-\ddx \left(\frac{1}{xu}\right) > 0\;.
\end{equation}

To proceed we need an expression for the spatial derivative of $u$:
\begin{equation}\label{eq-derivs}
\frac{du}{dx} = \left(\frac{P}{\rho^2}\frac{d\rho}{dt} - \frac{u}{\tau_c} + \frac{9}{4}\alpha c_s^2\Omega\right) \frac{dt}{dx} 
\end{equation}
where the time derivative of $\rho$ is given by
\begin{equation}\label{eqn-drhodt}
\frac{d\rho}{dt}= -3 \frac{\rho}{x}\frac{dx}{dt} \;.
\end{equation}
The coefficient 3 in Equation (\ref{eqn-drhodt}) arises from the spatial derivative of the enclosed mass and equals the dimension of collapse. 

The three terms in our expression for $du/dx$ (Equation \ref{eq-derivs}) come from $PdV$ work, cooling parametrized by the cooling time $\tau_c$ (c.f. Equation \ref{eqn-coolterm}), and viscous heating.  Balancing viscous heating against cooling gives rise to the Viscous Criterion.  In order to isolate the effect of $PdV$ work after fragmentation has occurred, we shall hereafter drop the last, viscous term, assuming that turbulent heating is inefficient within a fragment. This choice differentiates fragments from turbulent eddies.

For convenience we define the collapse timescale of the fragment:
\begin{equation}\label{eq-tcoll}
t_{ \rm coll} \equiv -\eqfrac{\rm{dlnx}}{\rm{dt}}^{-1} 
\end{equation}
Equation (\ref{eq-simpledx}) may now be written in the following simple form:
\begin{equation}\label{eq-stcool}
t_{\rm coll} \geq \tau_c (3\gamma - 4)
\end{equation} 
\Eq{eq-stcool} confirms that the object cannot collapse on a timescale faster than its own cooling time. We recover the standard result from stellar astrophysics that if $\gamma < 4/3$, thermal pressure cannot prevent collapse.  Also, an object cannot collapse on a timescale faster than its own free-fall time, 
\begin{equation}\label{eqn-ff}
t_{\rm ff} = \frac{1}{\sqrt{G\rho}} = \frac{\sqrt{2\pi Q}}{\Omega} \;,
\end{equation}
where to derive the final expression, we have assumed that $\rho = \Sigma/2H$ with $H=c_s / \Omega$ equal to the disk scale height.  Hence, the fragment collapses on whichever timescale is longer: $\tau_c(3\gamma - 4)$ or $t_{\rm ff}$.

Comparing these timescales, we find that free fall collapse requires:
\begin{equation} \label{eq-betalim}
\beta  \leq \frac{\sqrt{2\pi Q} }{(3\gamma-4)} \;.
\end{equation}
As expected, pressure is negligible when the cooling time is of order an orbital time. For $Q=1$ the limits on $\beta$ for free fall collapse are:
\begin{eqnarray}\label{eq-betaeval}
\beta \ltsim  2.5&,& \gamma = 5/3 \\ 
\beta \ltsim 13&,& \gamma = 7/5 \;.
\end{eqnarray}
These critical values of $\beta$ will be compared to the actual radiative cooling time of  fragments in different regimes in \S\ref{sec-irrad}. 

This calculation involves order unity coefficients which remain uncertain. Our primary goal is to establish the relevant physical parameters that must be evaluated to determine if fragments survive; in reality each value depends on the non-linear evolution which follows linear instability. High resolution hydrodynamic simulations will provide more insight than an analytic consideration of a range of fragment structures, densities and cooling laws. Note that our use of $\beta$ does not imply that the fragment cooling time must be equal to the background disk cooling time, although we expect the two to be quite similar just after the onset of fragmentation, which is the timescale addressed by this calculation.

\subsection{A Pressure Modified Hill Radius}\label{sec-KHcoll}
Now we consider the opposite regime, in which cooling is slow, and the fragment undergoes Kelvin-Helmholtz contraction rather than free-fall collapse.
 We calculate the critical size, $R_{\rm H+p}$, at which a pressure supported fragment can withstand tidal shear. In the absence of pressure support, this radius would just be the Hill radius (or Roche lobe).   The addition of pressure support is a small, though potentially significant, modification to this limit.  Pressure partially negates the force of self-gravity, leaving the object more susceptible to disruption by tidal forces.  

In binary stellar systems, modification of the Roche lobe due to pressure support is small because the scale height at the surface of the star is much smaller than the size of the star itself.  For fragments, this is not the case---the pressure scale length of the gas is comparable to the size of the fragment and pressure should be considered. Thus we are justified in our approximation of the pressure gradient here as $P/x$.
We calculate $R_{\rm H+p}$ for a compressible gas along the line of sight to the star, the direction in which tidal gravity is most disruptive.
 
Consider a fragment that has reached the Kelvin-Helmholtz contraction regime. The system is evolving on a cooling time, which we assume is substantially longer than a dynamical time.  This implies that self-gravity, pressure, and tidal forces are roughly in balance:
\begin{equation}\label{eqn-equilib}
\frac{G M_{ \rm enc} \rho}{x^2} - \frac{P}{x} - 3\Omega^2 x \rho= 0 \;.
\end{equation}
Because the fragment is contracting slowly, it remains in approximate equilibrium:
\begin{equation}\label{eq-dxbalance}
\ddx\left(\frac{G M_{\rm enc} \rho}{x^2} - \frac{P}{x} - 3\Omega^2 x \rho\right) = 0  \;, 
\end{equation}
and $x(t)$ is an undetermined function describing contraction. For simplicity, we rewrite this expression in terms of the dimensionless variable $B$:
\begin{equation}\label{eqn-slow}
\ddx\left[1 - B^{-1} - \eqfrac{x}{R_H}^3\right] = 0
\end{equation}
where $R_H = a\left({M_{\rm enc}/3M_*}\right)^{1/3}$ is the Hill Radius.

We plug Equations (\ref{eq-derivs}), (\ref{eqn-drhodt}), and (\ref{eqn-equilib}) into Equation (\ref{eqn-slow}) to calculate the contraction time as a function of $\tau_c$ and fragment size. Equation (\ref{eqn-slow}) reduces to:
\begin{equation}
\left(1-\eqfrac{x}{R_H}^3\right)\left(\frac{t_{\rm coll}}{\tau_c}-3\gamma +4\right)+3\eqfrac{x}{R_H}^3 =0
\end{equation}
Substituting in for different values of $\gamma$ we find:
\begin{eqnarray}\label{eq-fulltcool}
t_{\rm{coll}} =  \left\{\begin{array}{cl}
\displaystyle \tau_c \frac{4y-1}{y-1} \;\;\;& , \gamma = 5/3 \\
\rule{0ex}{5ex} \displaystyle \tau_c \frac{16y-1}{5y-5} &, \gamma = 7/5
\end{array}\right.
\end{eqnarray} 
where 
\begin{equation}
y=\eqfrac{x}{R_H}^3.
\end{equation}
 We can find the critical value of $y$ for which the clump will keep contracting by finding where $t_{\rm coll}$, given in Equation (\ref{eq-fulltcool}), changes sign: a negative contraction time signals expansion. This provides an estimate of the maximum size fragment that can be pressure supported but not torn apart by tides:
 \begin{eqnarray}\label{eq-xlimit}
{x} \leq R_{\rm H+p}= \eqfrac{1}{4}^{1/3} R_H, \gamma = 5/3 \\
{x} \leq  R_{\rm H+p}=  \eqfrac{1}{16}^{1/3} R_H , \gamma = 7/5 \label{eq-xlimit2}
\end{eqnarray}

\subsection{The Stalling Radius vs $R_{\rm H+p}$}\label{stalling-rad}
To interpret the severity of the hurdle embodied in \Eqs{eq-xlimit}{eq-xlimit2}, we must relate $R_{\rm H+p}$ to the initial size and mass of the fragment. A debate regarding the appropriate choice of these numbers is ongoing in the literature \citep[see, e.g.,][]{Boley2010}. The choices we make here are not the only sensible possibilities. We take the diameter of the collapsing region, $D_{\rm coll}$, to be half of the Toomre wavelength, $\lambda_T$, at $Q=Q_0\approx1$, envisioning that only half of the wavelength participates in forming an overdensity.  Since $\lambda_T = 2\pi H$, $D_{\rm coll} = \pi H$.  We expect this length scale to be at a minimum $2H$ for collapse to be roughly symmetric in the vertical and planar directions.\footnote{In reality, the scale height and thus wavelength may depend on the structure of the spiral arm in which the fragment forms \citep{Boley2010}.}  The radius of the collapsing region $r_{\rm coll} = D_{\rm coll}/2 = \lambda_T/4 = (\pi/2)H$, and the mass enclosed in the cylinder circumscribed by this radius is:
\begin{equation}
M_{\rm frag} = \Sigma_{Q=1} \pi (\lambda_T/4)^2.
\end{equation}

The Hill radius for a fragment with this mass is:
\begin{equation}
R_{H,\lambda_T} = \eqfrac{\pi^2}{12Q_0}^{1/3} H
\end{equation}
so that
\begin{equation}
\frac{R_{H,\lambda_T}}{\lambda_T/4} = \eqfrac{2}{3\pi Q_0}^{1/3} \approx 0.6
\end{equation}
Thus $\lambda_T/4$ cannot be  an estimate of the initial radial extent of the fragment for any extended length of time.  Rather, $\lambda_T/4$ likely represents the initial feeding zone for the fragment, not the fragment itself \citep{2005ApJ...621L..69R}. A collapsing fragment will only appear as a bound object when $x \lesssim R_H$.  
 
This interpretation is not an artifact of our choice of $r_{\rm coll}$. In a $Q=1$ disk, the only length scale on which one can create an object of mass $\Sigma \pi x^2$ that lives within its own Hill radius is smaller then the disk scale height:
\begin{equation}
R_{H, \rm{sc}} = \frac{H}{3Q}
\end{equation}
Creating such a fragment is inconsistent as it implies that the vertical and radial extent of the fragment differ by a factor of three.
Apparently, one cannot view fragments as chunks of the disk of radius $x$ extracted as with a cookie cutter from a sheet of dough. Perturbations must grow from the $Q=1$ state, and contract.  
 
Can such a contracting fragment achieve the radius required by Equation (\ref{eq-xlimit}) or (\ref{eq-xlimit2}) before pressure stalls collapse?  In general, answering this question requires following the non-linear evolution of the fragment, incorporating changes in the cooling time throughout this process.   Full consideration of this evolution is beyond the scope of the current work.  However, we obtain some guidance by considering the collapse of a fragment in the adiabatic (i.e. infinitely slow cooling) limit.  

Imagine that at the time of its initial linear collapse, a fragment has virial ratio $B = B_0 > 1$ but $-dB/dx < 0$ so that pressure is growing faster than self-gravity.  Assuming that rotational support does not become important (see Section \ref{caveats}), collapse will stall when $B \approx 1$.  The fragment will survive if tidal forces are not important at this stalling radius (i.e. if the stalling radius is smaller than $R_{\rm H+p}$).  For an adiabatic gas, $P \propto \rho^\gamma$, so 
\begin{equation}\label{eq-btox}
B \propto x^{(3\gamma-4)}
\end{equation}
 where we have assumed a constant $M_{\rm enc}$ so that $\rho \propto x^{-3}$.  Hence, the fragment stalls when its radius has been reduced by a factor of approximately $B_0^{-1}$ for $\gamma = 5/3$ or $B_0^{-5}$ for $\gamma = 7/5$.

For $r_{\rm coll} = \pi H/2$ and $Q_0=1$, $B_0 = \pi/2$. 
The fragment will initially start collapsing. In a $\gamma=5/3$ gas, the fragment only collapses to $\sim$60\% of its original size in the adiabatic limit (cooling would allow additional collapse). However, for $\gamma=7/5$, the collapse factor is nearly 10.  In fact, to contract from $\lambda_T/4 \sim 1.7 R_H$ to $0.4 R_H$ (c.f. Equation \ref{eq-xlimit2}) requires only a 
$\sim 34\%$ enhancement of gravity over pressure ($B_0\sim 1.34$)  
in the limit of infinite cooling time. This large factor suggests that even if the cooling time of the fragment becomes long, if its equation of state is soft, it can survive. This further suggests that fragment survival is trivial in disks that have both soft equations of state (including $\gamma = 7/5$, appropriate for molecular gas) and strong external irradiation. 

That the $\gamma=5/3$ case is less promising for fragmentation is neither surprising, nor as significant. We have only considered the adiabatic limit. Nevertheless we infer that fragmentation is much more likely to be inhibited by the Stalling Criterion for stiffer equations of state. { Note that the slow cooling limit differs from the adiabatic case in that the former is not dissipationless.  A truly adiabatic fragment could rebound after contraction.  We use this limit merely to show that slowly cooling fragments can contract below the stalling radius for soft equations of state.} 
Again, \Eq{eq-btox} reveals that fragmentation is difficult to inhibit in radiation pressure dominated disks with $\gamma=4/3$. This result is consistent with recent simulations by \cite{JiangGoodman11}.

\section{Other Disruption Mechanisms for Fragments} \label{caveats}
Thus far we have focused on thermal barriers to fragmentation. We now address dynamical and long timescale processes which may disrupt fragments. In particular we address the second classic cooling criterion mentioned in the introduction, the Collisional Criterion.

\subsection{The Collisional Criterion}\label{sec-ccrit}
 \cite{1989ApJ...341..685S} argue that slowly cooling fragments remain susceptible to disruption by neighboring fragments if the disk breaks up into objects separated by of order their own size, as expected from linear theory. Since the time to shear across a Hill radius at the Keplerian shearing velocity is of order an orbital period, fragments must contract on roughly the orbital timescale in order to avoid collisions.  For a $Q=1$ disk, the Keplerian shear energy at a separation of the Hill radius is of order the binding energy of a fragment, implying that encounters are likely disruptive.
More generally, \cite{1989ApJ...341..685S} suggest
\begin{equation} 
\beta \ltsim \frac{\Sigma_{\rm Q=1}} {\Sigma}\approx 1
\end{equation}
where $\Sigma_{\rm Q=1} $ is the critical surface density for fragmentation. This relation derives from the proportionality between $\Sigma$ and the wavelengths of growing modes in an unstable disk. Higher values of $\Sigma$ provide a wider range of growing modes. 

This argument ignores an important effect, which is that for soft equations of state (e.g. $\gamma = 7/5$), fragments with $\beta  > 1$ can still collapse on an orbital timescale. Moreover, for fragments stalled at $R_{\rm H+p}$, the binding energy should be larger than than the collisional energy by a factor of two, so fragments may avoid collisional disruption (and may not collide at all) for soft equations of state.

Global simulations suggest that  in systems with disk-to-central mass ratios $\gtsim 10\%$, fragments form in relative isolation within spiral arms \citep{LodRi05,Boley2007,KMKK10}, reducing the overall  frequency of collisions. Finally \cite{2004ApJ...608..108G} and \cite{2007MNRAS.374..515L} have argued that collisions might cause growth rather than disruption. These collisions have been observed in some global models \citep{Boley2010}.

\subsection{Rotation}
The most likely candidate for disruption that we have neglected is rotation.  Gauging the initial angular momentum of a fragment and predicting its evolution is challenging.  If angular momentum is conserved in a collapsing fragment, the object will eventually become rotationally supported. Dekel, et al., (in prep) report that fragments in isothermal disks collapse by a factor of $32/\pi \sim 10$ in radius before becoming rotationally supported . \cite{Boley2010} found that in SPH simulations with radiative cooling, the rotational support was only 20\% of the gravitational binding energy. 

In reality, subdisks eventually form around fragments, transporting away some fraction of the internal angular momentum. Since the disks are small, the viscous time is relatively short, about $\Omega_{\rm disk}^{-1}/\alpha_{\rm disk}$ allowing them to drain faster than the parent disk. Recent numerical work also suggests that gravitational torques can likely prevent the object from rotating faster than 50\% of break up \citep{Lin:2011}.  It is also possible that in the process of angular momentum transport, some mass is carried away. We make no attempt to account for this.

If the above estimates are correct, then rotational stalling likely occurs at radii smaller than $R_{\rm H+p}$, implying that rotational support does not inhibit fragmentation.  However, more numerical work is needed to confirm this conclusion.

\subsection{Migration}

Even if a fragment can survive at its current location, migration may bring it closer to the host star, and shrink $R_{\rm H+p}$ below the size of the fragment. 
 Because migration-induced disruption occurs on timescales longer than a few orbital times, even initially free-falling fragments might in principle suffer from late disruption due to this process, depending on the extent of the migration \citep{2005ApJ...633L.137V,Boley2010}.  
 
 \subsection{Interactions with Spiral Arms}

Close interactions with spiral arms generate a tidal force roughly equivalent to that from the central star, so collapsing in free fall or shrinking to $R_{\rm H+p}$ should protect fragments from this means of disruption. In the vicinity of an arm of width $\lambda_{\rm arm}$, the ratio of tidal gravity from the arm to tidal gravity from the star is approximately $(\Sigma_{\rm arm}\lambda_{\rm arm}^2/M_*)(r/\lambda_{\rm arm})^3 \sim (H_{\rm arm}/\lambda_{\rm arm})/(\pi Q_{\rm arm})$.

\section{Implications for Fragmentation}\label{sec-irrad}
We have argued that if a fragment can shed its angular momentum quickly enough to avoid rotational support, fragmentation may be generically possible in an irradiated disk with $\gamma = 7/5$, appropriate for the outer regions of protostellar disks.  This conclusion applies so long as $Q$ can reach its critical value, $Q_0 \sim 1$.  Here, we consider the conditions under which disks can be driven to $Q=Q_0$, while remaining irradiation dominated. We comment on the implications for fragmentation in protostellar disks (Section \ref{sec-protostellar}).  AGN disks (Section \ref{sec-AGN}) are less affected by these considerations.

\subsection{Protostellar Disks}\label{sec-protostellar}
We consider the case of a fragmenting disk surrounding an A-star. This choice is motivated by both theory and observations which suggest that A-stars may have companions consistent with formation by gravitational instability \citep{KMK08,KMCY10,2010ApJ...712..421H}. We calculate the equilibrium disk temperature for a $Q=Q_0 = 1$ disk, including both viscous heating and irradiation from the star.  

We begin by setting $\alpha = \alpha_{\rm sat}$, its saturation value for the gravitoturbulent regime, and we identify the irradiation-dominated regions within the disk given this choice.  Because $\alpha_{\rm sat}$ is not well-determined, we consider a range of possible values.  Because gravitoturbulence cannot support $\alpha > \alpha_{\rm sat}$, disk regions which are irradiation-dominated for $\alpha=\alpha_{\rm sat}$ will be irradiation-dominated in all $Q\sim 1$ disks.   
In contrast, regions of the disk which are viscously heated in this limit are not necessarily viscously heated at lower mass infall rates, $\dot M_{\rm inf}$, onto the disk, so after considering this limit, we turn to the lower values of $\alpha$ which can be generated by low $\dot M_{\rm inf}$.

We use the stellar irradiation model of  \cite{1997ApJ...490..368C}, which determines the incident flux on to the disk as a function of radius\footnote{ We note that this is not entirely self-consistent for a disk heated in part by accretion, but the effect is subtle.}. 
 The disk midplane temperature is found by solving the following equation:
\begin{equation}
\sigma T^4 = \frac{3}{8}\tau_R F_\nu + \sigma T_{\rm irrad}^4 
\end{equation}
where the energy flux due to viscous dissipation in the midplane is
\begin{equation}
F_\nu = \frac{3}{8\pi}{\Mdot \Omega^2}\;, 
\end{equation}
the mass accretion rate through the disk when $Q=1$ is $\Mdot \approx 3 \alpha c_s^3/G$, and
\begin{eqnarray}\label{eqn-TCG}
T_{\rm irrad}& = &\left[\frac{1}{(7\pi\sigma_B)^2}\frac{k}{\mu}\frac{1}{GM_*} \right]^{1/7}L_*^{2/7}r^{-3/7}\;.
\end{eqnarray}
In the following plots, we use
\begin{eqnarray}
T_{\rm irrad} &=& 40 K \eqfrac{r}{50\rm{AU}}^{-3/7} \eqfrac{L_*}{5 L_\odot}^{2/7}\eqfrac{M_*}{1.5\Msun}^{1/7}\label{eqn-AstarT} \,
\end{eqnarray}
where $\mu = 2.33 m_H$ and $m_H$ is the mass of a hydrogen atom.  From the tables accompanying \cite{Landin2009}, a pre-main-sequence A star has intrinsic luminosity $L_* \approx 2.5 L_\odot$ for ages of 1--3 Myr, with $L_\odot$ equal to the current luminosity of the Sun.  Infall is much more likely to drive disks unstable at earlier times, and at 3--8 $\times 10^5$ years, the intrinsic luminosity of a 1.5 $M_\odot$ star is approximately 7 $L_\odot$.  Accretion luminosity will exceed the intrinsic stellar luminosity for accretion rates onto the star $\gtrsim 10^{-7} M_\odot/$yr.  The stellar accretion rate may or may not match the mass infall rate $\dot M_{\rm inf}$.  We choose  $L_* = 5 L_\odot$ as an intermediate value. We use the cooling time defined in \Eq{eqn-tcooldef}.

To derive the most optimistic limits on fragmentation at small masses, we imagine that dust grains have grown to $\sim 300\mu$m 
so that the opacity $\kappa \approx 0.24$ cm$^2$g$^{-1}$ is constant with temperature (see \citealt{KMCY10} for a discussion of this choice and why it provides optimistic limits at small fragment masses). Given this disk model, we find the characteristic fragment mass $\Sigma \pi r_{\rm coll}^2$ for the disk where fragments can free fall collapse. Again to generate the smallest possible fragment masses, we use $r_{\rm coll }= \pi H/2$ (c.f. Section \ref{stalling-rad}, \citealt{Boley2010}).

\begin{figure}
\epsscale{1.0}
\plotone{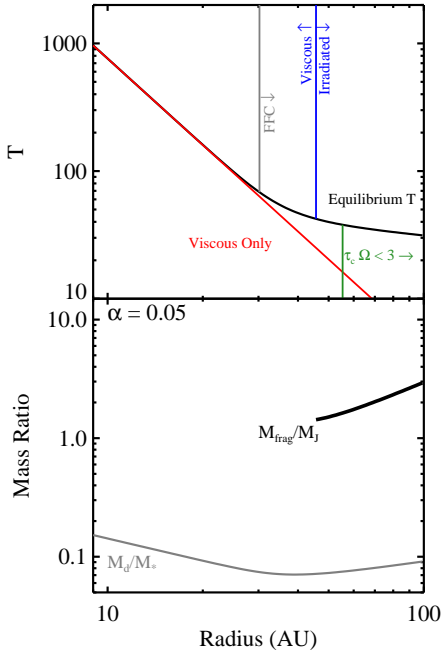}
\caption{Top: Equilibrium and viscous disk temperature for an A star as a function of radius for a disk with gravitoturbulent $\alpha = \alpha_{\rm sat} = 0.05$, for $L_* = 5 \Lsun$.  The vertical labelled lines show the oft used critical $\beta = 3$ (green), the dividing line between viscous, and irradiation dominated heating (blue), and the regime where free-fall collapse (FFC) is ensured  for $\gamma = 7/5$ (grey).  Bottom: corresponding minimum  initial fragment masses (black) where fragmentation is allowed and corresponding enclosed disk-star-mass ratio (grey) implied by the constant $Q$ model. Note that fragments will continue to grow---low initial masses do not guarantee the formation of planets over brown dwarfs.}
\label{fig-proto}
\end{figure}

\begin{figure}
\epsscale{1.0}
\plotone{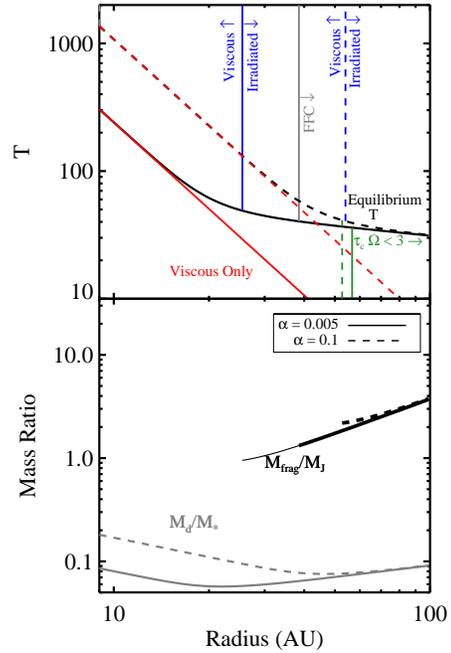}
\caption{ Similar to \Fig{fig-proto}, with temperatures, critical radii, and fragment masses shown for disks with $\alpha=\alpha_{\rm sat} = 0.005$ (solid) and $0.1$ (dashed).  Note there is only one line labelled FFC, corresponding to the $\alpha=0.005$ case because when $\alpha =0.1$, $\beta < 13.5$ at all radii shown.  In the large $\alpha$ case,  the viscous regime extends to larger distances. The thin portion of the fragment mass curve for $\alpha=0.005$ indicates the points at which free fall collapse is not guaranteed, but the disk remains irradiation dominated. Fragmentation in these locations is more uncertain. Again, these represent initial minimum masses; fragments will grow significantly towards the gap-opening and isolation mass \citep{KMCY10}}
\label{fig-both}
\end{figure}

 \Fig{fig-proto} shows the temperature profiles, critical radii where the cooling criteria are satisfied, and initial fragment masses for a disk with $\alpha_{\rm sat}=0.05$,  We show the corresponding temperatures, radii, and fragment mass scales for $\alpha_{\rm sat} = 0.005$ and $0.1$ in \Fig{fig-both}. For $\alpha_{\rm sat} > 0.05$, which is expected for GI-driven turbulence, the location at which the disk becomes irradiation dominated is effectively the same as the fragmentation boundary typically calculated from the Viscous Criterion. 
  
 For lower values of $\alpha_{\rm sat}$ fragmentation is pushed inwards, and there are disk locations that remain irradiation dominated, but where cooling prohibits free-fall collapse. Fragmentation is more uncertain in these regimes, but plausible if $\alpha_{\rm sat}$ ever takes such low values.

Recall that the critical value $\alpha_{\rm sat}$ { depends on the transport mechanism},  so in irradiation-dominated regions where global modes control transport,  the  saturation value for angular momentum transport will be different from the value set by gravitoturbulence and likely depends on disk mass \citep{Laughlin:1997fk}. Low mass, irradiation dominated disks with low infall rates  might be able reach $Q_0$ as discussed in Section \ref{sec-irraddrive}.  In this case, fragmentation could be pushed significantly closer to the star.

\begin{figure}
\plotone{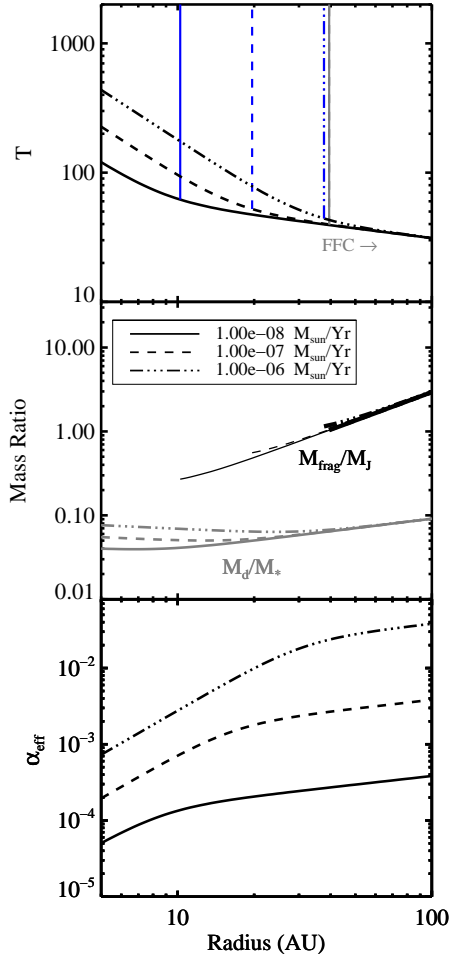}
\caption{Top: Black curves show the equilibrium disk temperature for three different infall rates, $\dot M_{\rm inf}$. The three blue, vertical lines indicate the radius outside of which the disks are irradiation dominated, while the grey vertical line (labeled FFC) shows where free fall collapse is ensured (it remains the same for all accretion rates). Middle: Disk and fragment masses as a function of radius. As in \Fig{fig-both}, the thin lines indicate where the disk is irradiation dominated but free fall collapse is inhibited. Bottom: Effective $\alpha$ values corresponding to the chosen temperature and accretion rate.  Even low values of $\alpha_{\rm eff}$ could exceed those for layered MRI driven accretion above a dead zone \citep{1996ApJ...457..355G}.}
\label{fig-allthatstuff}
\end{figure}

To determine whether this low disk mass, low accretion rate regime might be relevant, we calculate the equilibrium temperature for a $Q=1$ disk given a constant accretion rate (not fixed $\alpha_{\rm sat}$), and find the irradiation dominated regions of the disk. In \Fig{fig-allthatstuff} we show the temperatures, disk masses, fragment masses, and corresponding values of $\alpha$ for three different accretion rates. We must compare these with $\alpha_{\rm eff}$ generated by the MRI, as noted below. 

As discussed in Section \ref{sec-irraddrive}, $Q=Q_0$ is only likely to be achieved for irradiated disks with $M_d/M_* = H/r \lesssim 0.1$, evaluated at $r = r_{\rm out}$, where this critical value of  $0.1$ remains substantially uncertain.  
Using the \cite{1997ApJ...490..368C} irradiation temperature model,
\begin{eqnarray}
\frac{H}{r} &\approx& \left[\frac{1}{(7\pi\sigma_B)^2}\left(\frac{k}{\mu}\right)^8\right]^{1/14} \left(\frac{1}{GM_*}\right)^{4/7}L_*^{1/7} r^{2/7} \nonumber \\
&\approx& 0.07 \left(\frac{M_*}{M_\odot}\right)^{-4/7} \left(\frac{L_*}{5L_\odot}\right)^{1/7} \left(\frac{r_{\rm out}}{50\;{\rm AU}}\right)^{2/7} \label{eqn-Hoverr}
\end{eqnarray}
where we have assumed that the disk is optically thick at the outer edge when $Q \sim 1$.
From the pre-main-sequence evolutionary tracks of \cite{Landin2009}, stars with $M_* < 2M_\odot$ have intrinsic luminosities $L_* \approx 3.5 L_\odot (M_*/M_\odot)^{1.6}$ at an age of 3--8 $\times 10^5$ years and $L_* \approx 1.4 L_\odot (M_*/M_\odot)^{1.5}$ at 1--3 Myr.
These scalings suggest that at a given age, the value of $H/r$ decreases somewhat with increasing stellar mass, implying that spiral arm-driven transport may be more effective for disks around G stars than for those around A stars, particularly when low accretion rates do not generate significant stellar accretion luminosity.

\Fig{fig-allthatstuff} illustrates 
that disks can plausibly be low enough in mass to avoid global instabilities, but require more angular momentum transport than provided by the MRI.  The quantity $\alpha_{\rm eff}$ is averaged over the entire disk column---in dead zones experiencing only surface accretion, the $\alpha_{\rm eff}$ which can be supported by the MRI can be quite low \citep{1996ApJ...457..355G}.  In this case, the disk could be driven unstable while accreting at a lower rate than required in the viscous regime. For such low rates to drive fragmentation, the disk must already be near  $Q=1$ such that low accretion rates need not significantly alter the disk mass.

Since the exact criterion for global mode transport is unknown (see Section \ref{sec-irraddrive}), it is difficult to evaluate the importance of this regime.  We suggest that future simulations search for the maximum mass at which $Q\sim 1$, irradiated or isothermal disks do not show spiral arm-driven transport.

We have shown regimes in which planetary mass fragments might form and survive. We emphasize that newly born fragments must still be prevented from accreting surrounding disk material if they are to remain low in mass \citep{KMCY10}.  The isolation mass for such fragments remains in the brown-dwarf to stellar mass regime.

\subsection{AGN and Galactic Disks}\label{sec-AGN}
Contrary to the protostellar case, the question of interest in galactic scale disks is ``can fragmentation be prevented?" rather than ``can fragmentation be relevant for planets?" Relaxing the cooling criterion in the strongly irradiated portions of disks is of little import as these regions already satisfy the more restrictive Viscous Criterion for large $\alpha_{\rm sat}$, even at a fraction of a pc \citep{2003MNRAS.339..937G,2007MNRAS.374..515L}.  If galactic disks are irradiation dominated at radii as small  $0.01-0.1$ pc, fragments still collapse in free fall for modest black hole masses ($10^6\Msun$). The inner $0.01-0.1$ pc of hotter, more massive AGN disks ($M_{BH} = 10^7-10^8$)  are likely dominated by radiation pressure at these radii, shifting to $\gamma=4/3$, where free fall collapse in ensured. Recent numerical simulations by \cite{JiangGoodman11} are consistent with this analysis: they find fragmentation at long cooling times when radiation pressure dominates.

\section{Relevance of the Stalling Criterion in Viscously Heated Disks}\label{sec-visc}

In a recent suite of SPH simulations, \cite{MeruBate2010} find that the critical value of $\beta$ given by the Viscous Criterion increases with increasing resolution, with no convergence in sight. The authors suggest a possible breakdown of the Viscous Criterion. A complete breakdown of the Viscous Criterion would imply that dissipation and heating associated with mass transport occurs in an extremely non-local fashion.  While it seems impossible to avoid some critical $\beta$, its actual value  need not be near one, but rather is given by Equation (\ref{eq-tcoolalpha}), modulo small modifications due to spiral structure.

 It is dangerous to draw conclusions from un-converged simulations, but if the conversion between GI-driven turbulence and heating is treated properly in these runs, we suggest that rather than calling the Viscous Criterion into question, it implies that $\alpha_{\rm sat}$ is decreasing with increasing resolution, (a possibility that remains untested).  
One should not be surprised if  $\alpha_{\rm sat}$ remains un-converged at the maximum resolution, as the effective resolution is only marginally higher ($\delta x = 1/3H$) than that at which convergence was achieved in 2D \citep{Gam2001}.

\cite{LodatoClarke11} have suggested that deviations in $\alpha$ due explicitly to artificial viscosity are only a few percent.  However, it seems plausible that the resolution is simply too low to resolve the turbulence properly in the first place (another possibility which they acknowledge).  If the highest resolution calculations are converging on an accurate value of $\alpha_{\rm sat}$, it must be understood why it is a factor of $10-20$ smaller than that found by \citet{Gam2001}. 
The dimensionality of the calculations could explain the discrepancy;  similar critical $\beta$ values were found in 2D by \cite{Gam2001} and in low resolution SPH simulations which are effectively 2D due to a lack of scale height resolution \citep{2003MNRAS.339.1025R}.

If  the high resolution simulations have low  $\alpha_{\rm sat}$'s that are representative of the maximum transport provided by GI in low mass, thin disks, then this is worthy of further investigation. 
For sufficiently small $\alpha_{\rm sat}$, the Stalling Criterion may take over as the dominant cooling criterion.

\section{Summary and Discussion}\label{discussion}

In this paper, we separate the conditions for successful fragmentation into two considerations: (1) Can the disk reach the threshold of linear instability? and (2) Can a fragment survive after it has begun to collapse? This is distinct from previous work that has focused on whether or not the disk can reach $Q=1$ {\emph{ and}} cool quickly. We show that cooling is intimately connected both to reaching instability and to fragment survival.

We first address the distinction between disks heated internally by the dissipation of accretion energy, and those heated externally by irradiation. We show that the Viscous Criterion only governs the answer to question (1) in viscously heated disks because irradiated disks always effectively satisfy the Viscous Criterion.
We show that fragmentation is easier in irradiated disks, and that if an irradiated disk is thin enough (or equivalently low enough in mass) when $Q\sim1$, global transport may be ineffective, allowing the disk to be driven to collapse by low rates of mass infall.
Thus, fragmentation may occur at slightly smaller radii, and lower initial masses then previously assumed.  Importantly, fragmentation at low accretion rates could reduce the expected growth of fragments after their formation.

We then 
derive a generic cooling criterion which controls fragment survival in all disks. The Stalling Criterion divides fragments that can collapse on their own free-fall timescale from those that will become pressure supported and begin Kelvin-Helmholtz contracting. Fragments in the latter category must contract to a critical radius ($R_{\rm H+p}$) to avoid being torn apart by tides when pressure stalls their collapse. This radius is roughly half the Hill radius: the reduction is due to the inclusion of pressure support in the calculation of force balance.

For reasonable perturbations, fragments in disks with soft equations of state ($\gamma =7/5$) will likely be able to contract below $R_{\rm H+p}$, {\emph{even}} if they cool over many orbital periods.  In contrast, for stiffer equations of state, fragments that do not collapse in free fall will likely be sheared apart.

We note that the value of $R_{\rm H+p}$ depends on fragment geometry, and whether a given fragment will contract to this size depends on its non-linear evolution in time.
For example, if fragmentation occurs at densities and temperatures where there are sharp jumps in opacity, this must be taken into account in determing $\beta$.  Specific contraction models for disk-born fragments at all masses should be compared to the Stalling Criterion.  
We comment on other disruptive processes, including the classic Collisional Criterion, and the effects of fragment rotation, in Section \ref{caveats}.

Of course, since fragments will continue to grow following formation, gravitational instability should still favor the production of low-mass brown dwarfs over planets \citep{KMCY10}; That GI produces larger objects is increasingly supported by observations of M-star and brown-dwarf companions to A-stars \citep{2010ApJ...712..421H}, and the paucity of wide, planetary mass companions \citep{Leconte:2010}.

\subsection{Suggestions for Future Work}
This work should be a useful guide for future numerical simulations intended to study fragmentation in irradiation-dominated disks.  Two major uncertainties limit our conclusions for protostellar disks:
\begin{enumerate}
\item What is the value of $\alpha_{\rm sat}$ in a viscously-heated $Q\sim 1$ disk?
\item What is the maximum GI-driven transport rate as a function of $H/r$ in an irradiated $Q\sim1$ disk?  
\end{enumerate}
These two uncertainties are effectively the same question, but since the physics of transport in these regions differs, they must be considered independently.  Simulations will also be required to confirm our conclusion that regardless of whether they collapse in free fall, fragments can likely survive if $\gamma = 7/5$ because they collapse to smaller than $R_{\rm H+p}$ on of order a free-fall time. Hence, disks with soft equations of state must simply reach $Q=Q_0$ to fragment.  Confirmation of this conclusion will require a full treatment of radiative transfer and disk opacity.  As noted earlier, the impact of fragment rotation must also be explored.  
{ Current state of the art simulations of disks with radiative transfer do not yet cover the range of parameter space required to answer the above questions (\citealt{Boss:2007,2008ApJ...673.1138C, Cai:2010,Boley2010}; note that the latter two authors have questioned the conclusions of the former based on differing treatments of the photosphere).}

Answers from such simulations will enable us to distinguish between the following outcomes in each regime.
\begin{itemize}
\item {\it Viscously-heated regions:} If $\alpha_{\rm sat} \gtrsim 0.1$, then standard applications of the cooling time criterion give accurate results.  If $\alpha_{\rm sat} \lesssim 0.1$, then the standard Viscous Criterion will determine the boundary for fragmentation, but the critical cooling time will be longer.
\item {\it Irradiation-dominated regions:} If  both the maximum GI-driven transport rate and $\alpha_{\rm eff}$ due to MRI  are too low to process the infall rate onto a protostellar disk, fragmentation may be possible for thin disks at infall rates lower than $\sim$$c_s^3/G$.  If this occurs, low-mass fragments which do not experience substantial growth might be produced.  Otherwise, high accretion rates are required to drive these disks unstable. 
\end{itemize}

\acknowledgments{We thank the referee for detailed comments that improved this work. We also acknowledge Phil Armitage, Charles Gammie, Chris Matzner, and Ken Rice for useful conversations and for comments on an early version of this manuscript. KMK is supported by an Institute for Theory and Computation Postdoctoral Fellowship through the Harvard College Observatory.}

\appendix
\section{Behavior of $Q$ with $\Mdot$}
To determine under what conditions increasing $\Mdot$ drives disks more unstable, we consider the time evolution of $\Mdot$, $Q$, and $T$.
Reformulating \Eq{eqn-mdot}, the disk accretion rate becomes:
\begin{equation} \label{eqn-mdotalphaq}
\dot M = 3 \frac{\alpha(Q)}{Q}\frac{c_s^3}{G},
\end{equation}
where we have assumed that $\alpha$ is some unknown function of $Q$ \citep{1987MNRAS.225..607L, KMK08}. Because the evolution of $Q$ depends on how accretional heating is converted into midplane temperature, which is governed by the disk opacity, we must consider a temperature dependent opacity:
\begin{equation}
 \kappa =\kappa_0 T^{\beta_\kappa}. 
\end{equation} 
Solving for $Q/\alpha(Q)$ and using \Eq{eqn-accheat} to get $T$ as a function of $\Mdot$ we have:
 \begin{equation}\label{eqn-qoveralpha}
 \frac{Q}{\alpha} = \eqfrac{3}{G}^\frac{10-2\beta_\kappa}{7-2\beta_\kappa}\eqfrac{k_B}{\mu}^\frac{12-3\beta_\kappa}{7-2\beta_\kappa}\eqfrac{\kappa_0 \Omega^3}{16\pi^2\sigma}^\frac{3}{7-2\beta_\kappa}\frac{1}{\Mdot}\eqfrac{\Mdot}{Q}^\frac{3}{7-2\beta_\kappa} = a_1 \frac{1}{\Mdot}\eqfrac{\Mdot}{Q}^\frac{3}{7-2\beta_\kappa},
 \end{equation}
where we have rewritten $\Sigma$ using the definition of $Q$.  We subsume the first three terms of \Eq{eqn-qoveralpha} into  a constant $a_1$ for convenience.  
Rearranging, we find:
 \begin{equation} \label{eqn-Qbeta}
 Q = a_1^\frac{7-2\beta_\kappa}{10-2\beta_\kappa}\alpha^\frac{7-2\beta_\kappa}{10-2\beta_\kappa}\Mdot^\frac{\beta_\kappa-2}{5-\beta_\kappa}
 \end{equation}
 Because $\alpha$ is an unknown function of $Q$, we relate the time derivatives of $\alpha$ and $Q$ using an unknown function $C_1(Q) > 0$:
\begin{equation}
\frac{1}{\alpha}\frac{d\alpha}{dt} = -C_1 \frac{1}{Q}\frac{dQ}{dt}.
\end{equation}
This relation captures the fact that we expect $\alpha$ to increase as $Q$ declines, and allows us to treat the above expression analytically.  Taking the time derivative of \Eq{eqn-Qbeta} we find:
\begin{equation}\label{eqn-betafrac}
\left(1+\frac{7-2\beta_\kappa}{10-2\beta_\kappa}C_1\right)\frac{1}{Q}\frac{dQ}{dt} = \frac{\beta_\kappa-2}{5-\beta_\kappa}\frac{1}{\Mdot}\frac{d\Mdot}{dt}.
\end{equation}
We can immediately see that the sign of the relation between the derivative of $\Mdot$ and the derivative of $Q$ is governed by the value of $\beta_\kappa$, and in some cases $C_1$. To aid in our interpretation of \Eq{eqn-betafrac}, we can also relate $dT/dt$ and $d\Mdot/dt$:
\begin{equation}\label{eqn-Tmdotfrac}
\frac{1}{T}\frac{dT}{dt} = \left(2+\frac{C_1(\beta-2)}{5-\beta +C_1(7/2-\beta)}\right)\frac{1}{5-\beta}\frac{1}{\Mdot}\frac{d\Mdot}{dt}
\end{equation}
Together these two equations delineate four different $\beta_\kappa$ regimes:
\begin{enumerate}
\item{$\beta_\kappa < 2$: In this case, the right hand side of \Eq{eqn-betafrac} is negative, so that higher accretion rates drive the disk more unstable. Thus for small $\beta_\kappa$, an increase in $\Mdot$ causes a modest increase in temperature and a larger increase in $\Sigma$, causing $Q$ to decline.  This case is most relevant to protostellar disks. See \cite{ML2005}  for a discussion of the $\beta_\kappa = 2$ case.}
\item{$2<\beta_\kappa<7/2$: Both sides of \Eq{eqn-betafrac} are positive, meaning that an increase in $\Mdot$ actually drives $Q$ up. This occurs because the steep increase in opacity with $T$ causes the temperature to rise faster than $\Sigma$, increasing $Q$.}
 \item{$\beta_\kappa > 5$: When $\beta_\kappa$ is extremely high, an increase in $\Mdot$ always corresponds to a decrease in $Q$.  This dependence derives from the fact that for $\beta_\kappa >5$,  an equilibrium between accretional heating and cooling as defined by \Eq{eqn-accheat} requires that an increase in $\Mdot$ correspond to a {\emph{decrease}} in temperature, which is counterintuitive ($T \propto \Mdot^\frac{2}{5-\beta_\kappa}$ at constant $\alpha$).  }
\item{$7/2<\beta_\kappa<5$: The sign of the left hand side of \Eq{eqn-betafrac} depends on the value of $C_1$, and thus the functional form of $\alpha(Q)$. Increasing $C_1$ shifts the value of $\beta$ at which an increase in $\Mdot$ corresponds to a decrease in $T$, as shown in \Eq{eqn-Tmdotfrac}. When $C_1\rightarrow 0$, the disk behaves as in case 2 all the way to $\beta =5$.   When $C_1$ is large,  this case behaves as case 3, with the upper limit on $\beta$ a function of $C_1$.}
\end{enumerate}

The outer regions of protostellar disks on which we focus here are all encompassed by case 1, in which increased accretion destabilizes the disk.

%\bibliography{diskbibU}

\end{document}